\documentclass{ws-rv9x6}
\usepackage{subfigure}     
\usepackage{ws-rv-van}     
\makeindex
\begin{document}

\chapter[Using World Scientific's Review Volume Document Style]{Berry phase and backbending}\label{ra_ch1}

\author[P. Ring]{P. Ring\footnote{ring@ph.tum.de}}

\address{Physikdepartment, Technische Universit\"at M\"unchen,\\
85748 Garching, Germany}

\begin{abstract}
Backbending is a typical phenomenon in the rotational spectra
of superfluid nuclei. It is caused by the rotational alignment
of a pair of nucleons and depends on topological properties of
the Hartree-Fock-Bogoliubov spectrum in the rotating frame
characterized by diabolic points and Berry phases.
\end{abstract}

\body

\section{Introduction}
\label{sec1}

Nuclei are finite strongly interacting many-body systems. Because of the Pauli principle and the uncertainty relation, nucleons in the medium only feel a reduced effective interaction, such that the mean field approach can be applied with considerable success. This is, in particular, true for nuclei far from closed shells where one observes phase transitions and where correlations can be taken into account in an effective way through the spontaneous violation of symmetries such as the rotational symmetry in deformed nuclei and the gauge symmetry in superfluid systems. As a result, the wave functions can be described within deformed Hartree-Fock-Bogoliubov (HFB) theory~\cite{Ring1980} in an intrinsic system by simple generalized product states $\vert\Phi\rangle$.

This simplicity not only allows us a microscopic description of the nuclear ground state, but HFB theory in the rotating frame is also a very powerful tool for investigating complicated processes in high spin physics: collective phenomena such as shape changes and phase transitions, as well as single particle properties such as rotational alignment processes.

\begin{figure}[th]
\centerline{\psfig{file=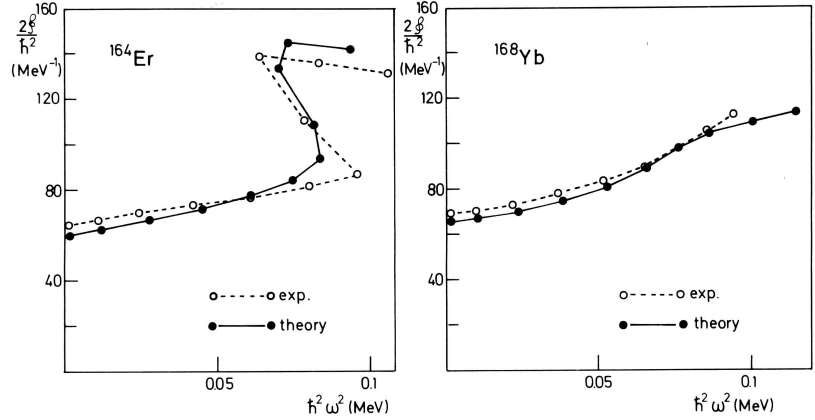,width=11.0cm}}
\caption{Backbending plots in Rare Earth {nuclei\cite{Mang1976_ZPA279-325}}. The experimental values for $\omega$ and $\cal J$ are obtained from Eqs. (\ref{frequency}) and (\ref{inertia}) by discretization.}
\label{fig1}
\end{figure}

In this Chapter we concentrate on the backbending phenomenon found in many of the rotational bands of well deformed superfluid nuclei. It leads to steeply increasing and even backwards bending curves in plots of the moment of inertia as a function of the angular velocity (see Fig.~\ref{fig1}). We identify this as a level-crossing phenomenon between the ground state band and a band of two aligning quasiparticles. It is characterized by topological properties of the rotating HFB spectra and we shall discuss consequences for the oscillating behavior of backbending and diabolic pair transfer.

\section{Cranked Hartree-Fock-Bogoliubov theory}
\label{sec2}

The exact many-body wavefunctions $\vert\Psi\rangle$ of the finite nuclear system have to obey the full symmetries of the Hamiltonian, i.e. they are eigenstates of particle number and angular momentum
with quantum numbers $N$ and $I$ respectively. Since the intrinsic wave functions $\vert\Phi\rangle$ break these symmetries the transformation from the intrinsic frame to the laboratory system is connected through a projection to good quantum numbers.
\begin{equation}
\vert\Psi^{NI}\rangle = \hat{P}^N\hat{P}^I\vert\Phi\rangle
\end{equation}
The {\it yrast} states are the lowest states for given $N$ and $I$ and therefore their intrinsic functions $\vert\Phi\rangle$ have to be determined by variation after projection\cite{Zeh1965_ZP188-361,Egido1982_NPA383-189}
\begin{equation}
\delta E^{NI}=\delta\frac{\langle\Phi\vert\hat{H}\hat{P}^N\hat{P}^I\vert\Phi\rangle}{\langle\Phi\vert\hat{P}^N\hat{P}^I\vert\Phi\rangle}=0,
\end{equation}
Technically the projection is a complex mathematical process, but for large particle numbers and strong deformations with a well defined orientation it can be shown\cite{Beck1970_ZP231-26,Mang1975_PR18-325} that the intrinsic wave function is to a good approximation given by a Cranking wave function $\vert\Phi(\omega)\rangle$ deduced from a variation of the unprojected Hamiltonian in a system rotating with constant frequency $\omega$
\begin{equation}
\langle\delta\Phi\vert\hat{H}-\lambda\hat{N}-\omega\hat{J}_x\vert\Phi\rangle=0.
\end{equation}
Angular momentum and particle number are only conserved on the average. $\omega$ and $\lambda$ are determined by the semiclassical
condition introduced by Inglis\cite{Inglis1956_PR103-1786}
\begin{equation}
J:=\langle\Phi \vert \hat{J}_x \vert \Phi\rangle = \sqrt{I(I+1)}~~~~~{\rm and}~~~~~\langle\Phi \vert \hat{N} \vert \Phi\rangle = N.
\label{Inglis}
\end{equation}
This leads to the HFB equations in the rotating {frame\cite{Ring1970_ZP231-10}}. Strongly deformed axially symmetric nuclei rotate around an axis ($x$) perpendicular to the symmetry axis ($z$). Time reversal symmetry is broken in the rotating frame, but the system is invariant with respect to a rotation of 180$^\circ$ around the $x$-axis with the signature quantum number $\pm i$. In this case these equations are real and can be reduced to an equation of half the number of dimensions
\begin{equation}
\left(
\begin{array}
[c]{cc}%
h-\lambda-\omega j_{x} & \Delta\\
\Delta & -(h-\lambda+\omega j_{x})
\end{array}
\right)
\left(
\begin{array}
[c]{c}%
U\\
V
\end{array}
\right)  _{k}=E_{k}\left(
\begin{array}
[c]{c}%
U\\
V
\end{array}
\right)_{k}
\label{HFB-rot}
\end{equation}
with a similar equation for the other signature. $h=t+\Gamma$ contains the kinetic energy $t$ and the mean field $\Gamma(\rho)$ and $\Delta(\kappa)$ is the pairing field. The fields depend in a self-consistent way on the density matrix $\rho=\langle\Phi\vert a^\dag a\vert\Phi\rangle$ and the pairing tensor $\kappa=\langle\Phi\vert a^\dag a^\dag\vert\Phi\rangle$.
$E_k(\omega)$ are the quasiparticle energies in the rotating frame (Routhians) and $\vert\Phi\rangle$ is formed as a generalized Slater determinant of the quasiparticle wave functions $U_k$ and $V_k$. The upper part of Eq.~(\ref{HFB-rot}) corresponds to basis vectors with signature $+i$, the lower part to signature $-i$ with the pairing field $\Delta$ connecting both signatures.

The solution of these self-consistent equations allows one to calculate the rotational spectrum $E(\omega)=\langle\Phi\vert\hat{H}\vert\Phi\rangle$ as a function of the angular velocity or, using the Inglis condition (\ref{Inglis}), as a function of $J$ and we find
\begin{equation}
\omega = \frac{dE}{d J}~~~~~~~~{\rm and}~~~~~~~~J = -\frac{dE^\prime}{d\omega},
\label{frequency}
\end{equation}
where $E^\prime=E-\omega J$. One can define two moments of inertia, either the {\it static} moment ${\cal J}^{(1)}$ or the {\it dynamic} moment ${\cal J}^{(2)}$, where
\begin{equation}
{\cal J}^{(1)} = \frac{J}{\omega}=\left(\frac{1}{J}\frac{d E}{dJ}\right)^{-1}~~~~{\rm and}~~~~~~~{\cal J}^{(2)} = \frac{dJ}{d\omega}=\left(\frac{d^2 E}{d J^2}\right)^{-1}.
\label{inertia}
\end{equation}
The self-consistent solution of the rotating HFB equations (\ref{HFB-rot}) for even~\cite{Ring1970_ZP231-10} and for odd~\cite{Ring1974_NPA225-141} particle numbers allows a microscopic description of the complicated interplay between the collective rotational and single particle degrees of freedom, including changes of the shape (stretching) and of the pairing field up to the point of pairing collapse.

A simplified version of cranked HFB theory with constant fields $\Gamma$ and $\Delta$ has been introduced by Ring and {Mang\cite{Ring1974_PRL33-1174}}. It has been used with great success under the name {\it cranked shell model} (CSM) by Bengtsson and Frauendorf\cite{Bengtsson1979_NPA327-139} to analyze rotating quasiparticle spectra.

\section{Backbending as a level crossing phenomenon}
\label{sec3}
At the end of the sixties heavy ion fusion reactions allowed one to populate nuclear states with high angular momentum and to observe the $\gamma$-decay along the {\it yrast line} and to   analyze these rotational spectra in terms of angular velocities and moments of inertia. For a rigid rotor one would have expected a constant value for the moment of inertia. Non-linear effects, such as stretching or a reduction of pairing were expected to show deviations proportional to $\omega^2$, but in 1971 a new and unexpected phenomenon was discovered by Johnson, Ryde, and {Sztarkier\cite{Johnson1971_PLB34-605}}, {\it backbending}. For some nuclei (see Fig.~\ref{fig1}), at a certain critical angular velocity $\omega_c$ the curve ${\cal J}^{(2)}$ increased steeply and even bent backwards as a function of $\omega^2$. Initially the physical origin of this behavior was not at all clear. It could have been a sudden change in shape, a pairing collapse as predicted by Mottelson and Valatin\cite{Mottelson1960_PRL5-511} or, as proposed by Stephens and Simon\cite{Stephens1972_NPA183-257} a crossing of the ground state band (g-band) with a highly aligned two-quasiparticle band (s-band), in which the two particles occupy high-$j$ intruder levels, in this case neutron $1i_{13/2}$ levels. The self-consistent solution of the cranked HFB equations by Banerjee, Mang and Ring\cite{Banerjee1973_NPA215-366} showed that the latter was the right solution. Backbending is caused by alignment. The deformation stays roughly constant, but the pairing correlations are somewhat reduced because of blocking of the two quasiparticles.

\begin{figure}[th]
\centerline{\psfig{file=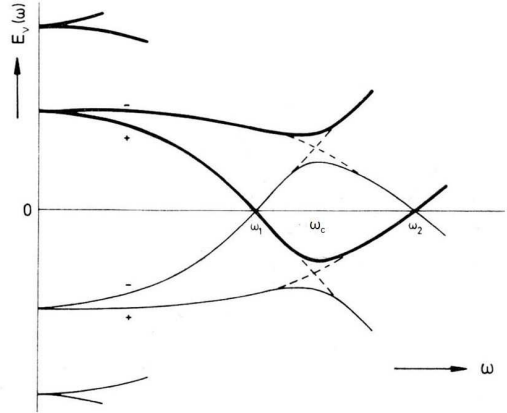,width=7.0cm}}
\caption{Schematic quasiparticle spectrum\cite{Banerjee1973_NPA215-366} of the rotating HFB equations (\ref{HFB-rot}). Bold curves indicate the occupied levels  and +/- give their signature $\pm i$.
\label{fig2}}
\end{figure}

In order to discuss this level crossing in more detail, we have to understand the special structure of the quasiparticle spectrum. It is shown schematically in Fig.~\ref{fig2} for
constant fields $\Gamma$ and $\Delta$. The eigenvalues of the HFB equations form in pairs $\pm E_k$. For each $k$ one has to choose one of these vectors. In the simple Hartree-Fock case, this choice corresponds to the freedom to occupy a level $k$ or to leave it empty. In the ground state one fills the levels from the bottom of the well and this corresponds in the HFB solution to choosing only positive values of $E_k>0$. For $\omega=0$ one has in even-even systems time-reversal invariance and therefore the eigenvalues $E_k$ for the two signatures are two-fold degenerate and they show a gap typical for the BCS eigenvalues $E_k=\sqrt{(\varepsilon_k-\lambda)^2+\Delta^2}$.

For increasing frequencies $\omega$ we find signature splitting. In some spectra (as in Fig.~\ref{fig2}) the lowest eigenvalue goes to zero ({\it gapless superconductivity}), where it crosses sharply the corresponding negative eigenvalue having different signature. At somewhat higher frequencies $\omega_c$ we observe an avoided crossing between two levels with the same signature. This corresponds to a crossing of the g-band with the $s$-band of two strongly aligned $\nu 1i_{13/2}$ quasiparticles. It leads to the sharp increase of the moment of inertia shown in Fig.~\ref{fig1}. Of course, the backward going part of this curve cannot be explained with for fixed fields $\Gamma$ and $\Delta$. The theoretical results in Fig.~\ref{fig1} have been obtained for a self-consistent solution with changing fields using the gradient method\cite{Mang1976_ZPA279-325} constraining the angular momentum in Eq.~(\ref{Inglis}), because, for constant values of $\omega$, the solution on the backward going branch does not correspond to a minimum in the energy {surface\cite{Chu1975_PRC12-1017}}.

In fact, a sharp level crossing of the ground state band $\vert\Phi_0\rangle$ and a two-quasiparticle band $\alpha^\dag_1\alpha^\dag_2\vert\Phi_0\rangle$ with very different angular momenta cannot be described properly at constant angular velocity $\omega$ using a single Slater determinant $\vert\Phi(\omega)\rangle$. In reality the mixing occurs at the same angular momentum $I$ with wavefunctions belonging to different angular velocities. In a single $j$ model with fixed deformation and pairing, Hamamoto\cite{Hamamoto1976_NPA271-15} could show that the cranking model introduces a spurious mixing with an exaggerated interaction between the two levels in the region of the crossing in Fig.~\ref{fig2}.
To avoid this problem, in practical CSM applications\cite{Bengtsson1979_Nucl.Phys.A314-27} the two bands are often connected in an adiabatic way by dashed lines. Technically this is not always simple and the proper microscopic solution of this level crossing problem is given by a mixing at constant angular momentum in the projected shell model (PSM) of Hara and Sun\cite{Hara1995_IJMPE4-637} where the wavefunction is given by a projected linear combination of a deformed ground state $\vert\Phi_0\rangle$ and many two-quasiparticle states
\begin{equation}
\vert\Psi^{NI}\rangle = \hat{P}^{N}\hat{P}^{I}\{c_0\vert\Phi_0\rangle + \sum_{KK'}c_{KK'}\alpha^\dag_K\alpha^\dag_{K'}\vert\Phi_0\rangle\}.
\label{PSM}
\end{equation}
The coefficients $c_0$ and $c_{KK'}$ are determined by variation after projection.
Unfortunately this method is numerically rather involved. There are, however, many successful applications based on simple separable {models\cite{Hara1995_IJMPE4-637}}.

Summarizing this section we can say that the backbending phenomenon in rotational bands of deformed nuclei presents a rather complicated interplay of collective and single particle degrees of freedom. The Coriolis force tries to align the single particle angular momenta in the direction of the rotational axis. It counteracts the pairing forces favoring pairs with opposite angular momenta. The resulting Coriolis anti pairing effect, however, does not cause a collective alignment with pairing collapse, as one finds it in simple degenerate  {models\cite{Chu1975_PRC12-1017}}.  High-$j$ intruder orbits feel a stronger Coriolis force and align first and one observes a band crossing between the ground state and the lowest aligned band.

If the residual interaction between these two bands is small one has a sudden transition and backbending, otherwise a smooth alignment with a slowly increasing moment of inertia. Because of self-consistency the collective properties of the nucleus are also involved in this {transition\cite{Banerjee1973_NPA215-366}}. In the aligned band one of the pairs does not participate in pairing and therefore, because of the blocking effect, pairing is reduced by roughly 20 \%. The alignment also causes small triaxial $\gamma$-deformations. It requires, however, rather high spin to align more and more pairs in order to generate substantial triaxiality and only for the terminating {bands\cite{Afanasjev1999_PR322-1}}, where all the particles in the valence shell are aligned, does one find transitions to oblate shapes.

\section{Diabolic points in cranked HFB spectra}
\label{sec4}

Over the years, many rotational bands have been found, but as indicated in Fig.~\ref{fig1} not all of them show backbending. From the previous considerations it is clear that backbending occurs only, if the interaction between the two crossing bands is weak enough. For strong interactions one observes a smooth increase of the moment of inertia.

\begin{figure}[t]
\centerline{
\psfig{file=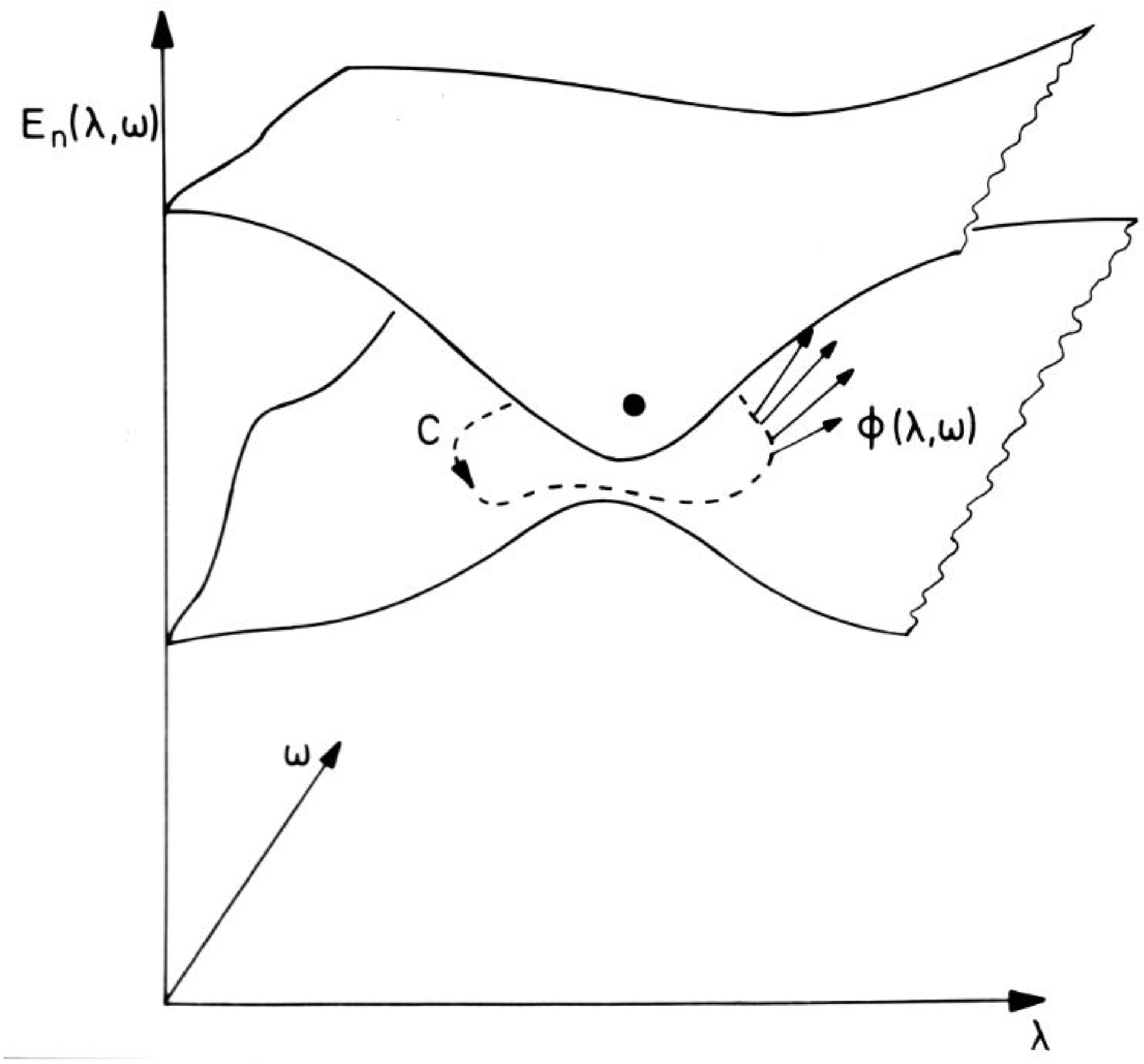,width=6.0cm}
~~~~
\psfig{file=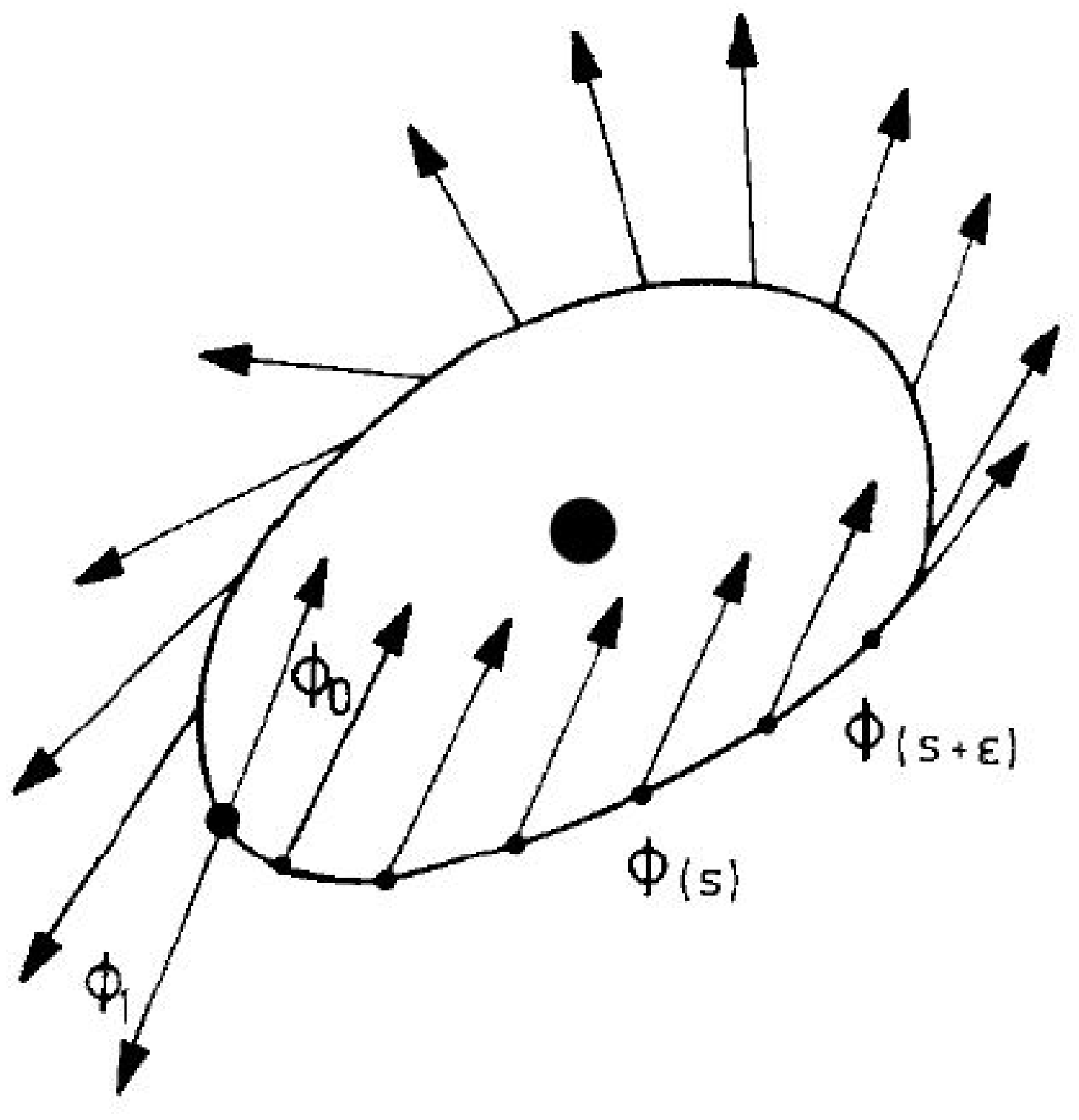,width=3.8cm}}
\caption{(Left panel) schematic representation~\cite{Ring1987_JPC48-339} of a diabolic point, where two energy surfaces touch.
A closed path $\cal C$ is shown encircling this point. (Right panel) the line bundle of wave functions
on a closed path producing a Berry phase -1: $\phi_1=-\phi_0$.
\label{fig3}}
\end{figure}

Of course, the phenomenon would be most pronounced for vanishing interaction, i.e. for a sharp crossing of the two levels with the same signature as in Fig.~\ref{fig2}. According to the {\it no-crossing rule} of von Neumann and Wigner\cite{Neumann1929_ZP30-427} two eigenvalues with the same symmetry should never cross. This is, however, not a theorem but only a rule. There are exceptional points, in particular for Hamiltonians depending on two parameters as shown schematically in Fig.~\ref{fig3}. Each eigenvalue $E_n$ forms a surface and there are {\it diabolic points}, where two of these surfaces touch. In the neighborhood of such a point, the two touching surfaces have the structure of a double cone, which reminded Berry\cite{Berry1984_PRSocA392-15} of an Italian toy called a "diabolo".

\begin{figure}[th]
\centerline{
\psfig{file=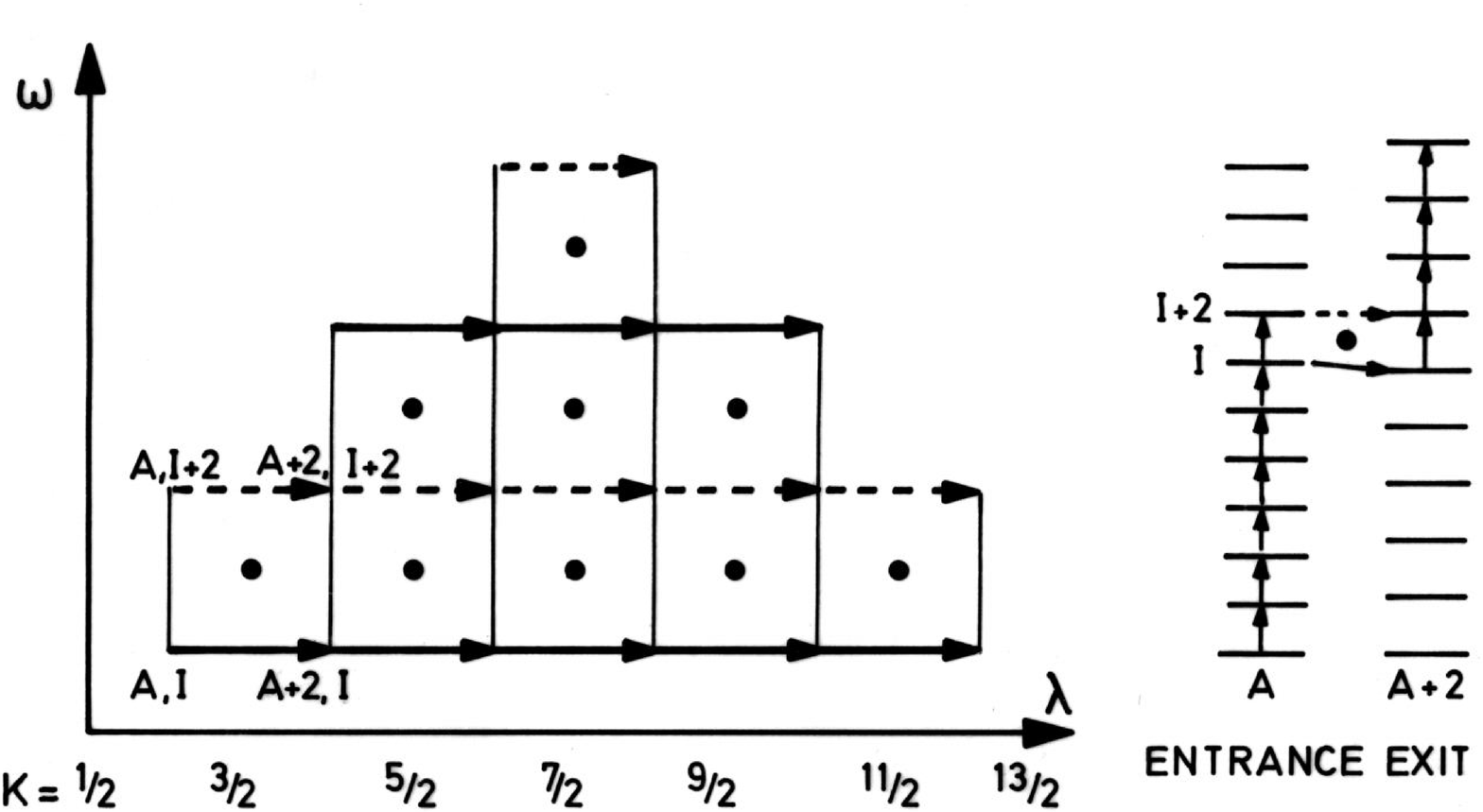,width=10.0cm}}
\caption{Diabolic point connecting two quasiparticle energy surfaces in the $(\lambda,\omega)$-plane.
\label{fig4}}
\end{figure}

Long before Berry had introduced this nomenclature, Bengtsson, Hamamoto and Mottelson\cite{Bengtsson1978_PLB73-259} had, in 1978, found that there are such exceptional points in the spectrum of the rotating HFB Hamiltonian (\ref{HFB-rot}) which depended on two parameters $\lambda$ and $\omega$. As shown by Nikam and Ring\cite{Nikam1987_PRL58-980} they are distributed in the ($\lambda,\omega$)-plan according to a very characteristic pattern, which is determined by the single particle angular momentum $j$ of the intruder shell. In the Rare Earths this is the $\nu 1i_{l3/2}$-orbit. The corresponding pattern is shown schematically in Fig.~\ref{fig4}. In the nuclear potential of prolate shape this intruder orbit with positive parity is split and the two-fold degenerate sub-levels having the magnetic quantum numbers $\pm K$ are distributed between the remaining neutron levels of different parity. With increasing neutron number, i.e. with growing chemical potential $\lambda$, these levels are filled pairwise according to their energy in the order $K=\pm\frac{1}{2},\pm\frac{3}{2}\dots\pm\frac{13}{2}$. We observe sharp backbending close to the diabolic points, where the interaction between the levels is small, and a smooth increase of the moment of inertia in the area far from these points. This leads to the {\it oscillating behavior of backbending\cite{Bengtsson1978_PLB73-259,Grummer1979_NPA326-1}}.

Beside the row of 5 diabolic points having small $\omega$-values we have further diabolic points so that we can expect further backbending phenomena at higher spin values.
In order to understand the pattern, at least in a qualitative way, we consider a single-$j$ shell embedded in a potential with deformation $\kappa$ and a constant pairing field with gap parameter $\Delta$. Writing this Hamiltonian in the basis $\vert\varphi^{(+)}_\nu\rangle_\omega$ and $\vert\varphi^{(-)}_{\bar{\nu}}\rangle_\omega$ of the eigenfunctions of $h-\lambda\pm\omega j_x$ with eigenvalues $\varepsilon^{(\pm)}(\omega)$  we end up with the matrix
\begin{equation}
\left(
\begin{array}
[c]{cc}%
\varepsilon^{(+)}(\omega) & \Delta(\omega)\\
\Delta(\omega) & -\varepsilon^{(-)}(\omega)
\end{array}
\right)
\label{HFB-rotbas}
\end{equation}
having diagonal parts $\varepsilon^{(+)}_\nu$ and $\varepsilon^{(-)}_{\bar\nu}$  and an off diagonal pairing field $\Delta_{\nu\bar{\nu}}(\omega)$. For small values of the parameter $\Delta$ perturbation theory applies and only the diagonal matrix elements are important
\begin{equation}
\Delta_{\nu\bar{\nu}}(\omega)=\Delta\langle\varphi^{(+)}_{\nu}\vert\varphi^{(-)}_{\bar{\nu}}\rangle_\omega=
\Delta\langle A+2\vert a^\dag_{\nu} a^\dag_{\bar{\nu}}\vert A \rangle_\omega
\label{overlap}
\end{equation}
These are proportional to the spatial overlap $\langle\varphi^{(+)}_{\nu}\vert\varphi^{(-)}_{\bar{\nu}}\rangle_\omega$ of eigenfunctions of opposite signature within the deformed rotating field. These integrals are identical to the pair transfer matrix elements between two unpaired nuclei with particle numbers $A$ and $A+2$, in which the two additional nucleons sit in the rotating orbits having quantum numbers $\nu$ and $\bar{\nu}$.

\begin{figure}[tbh]
\centerline{
\psfig{file=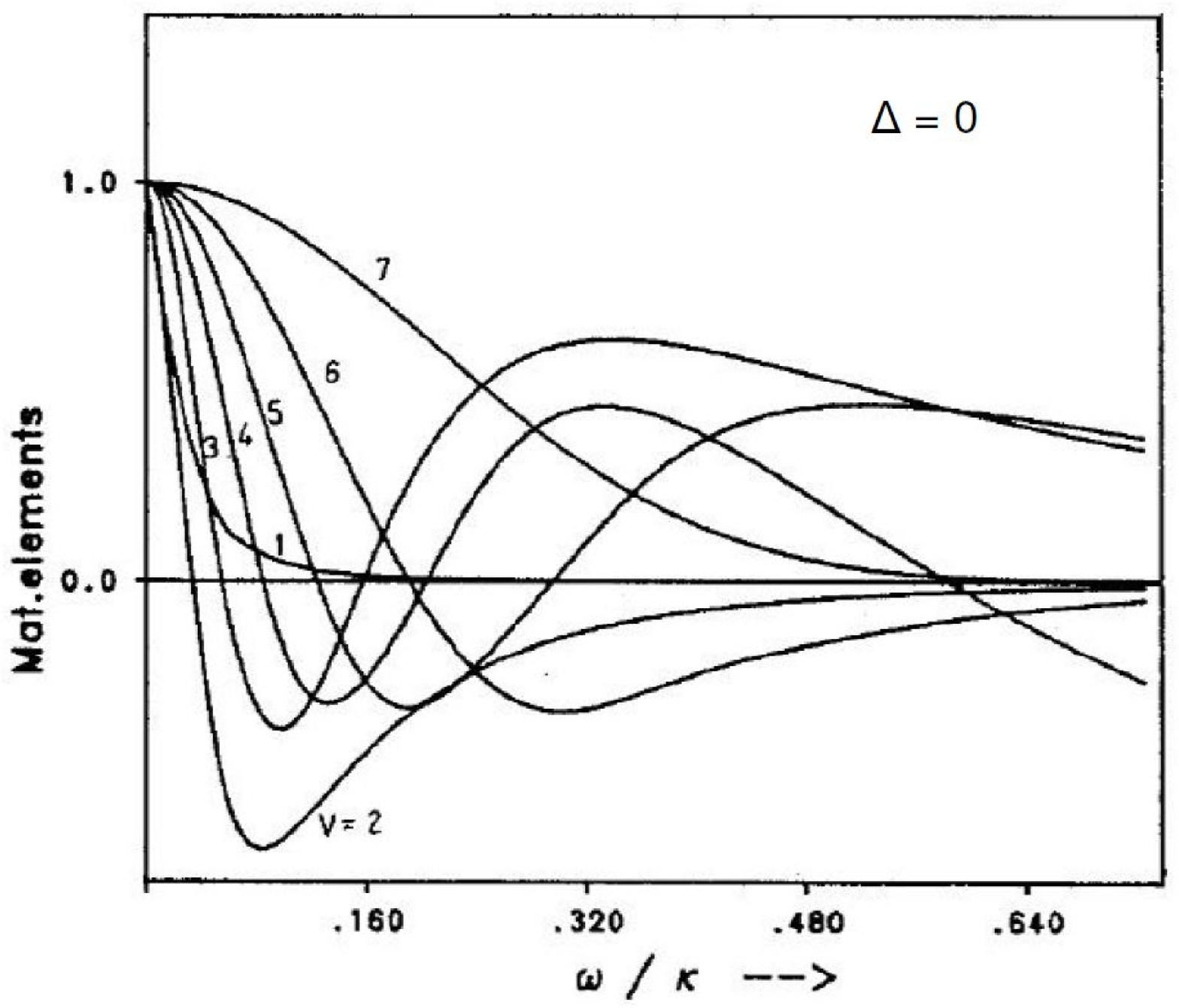,width=5.0cm}
\psfig{file=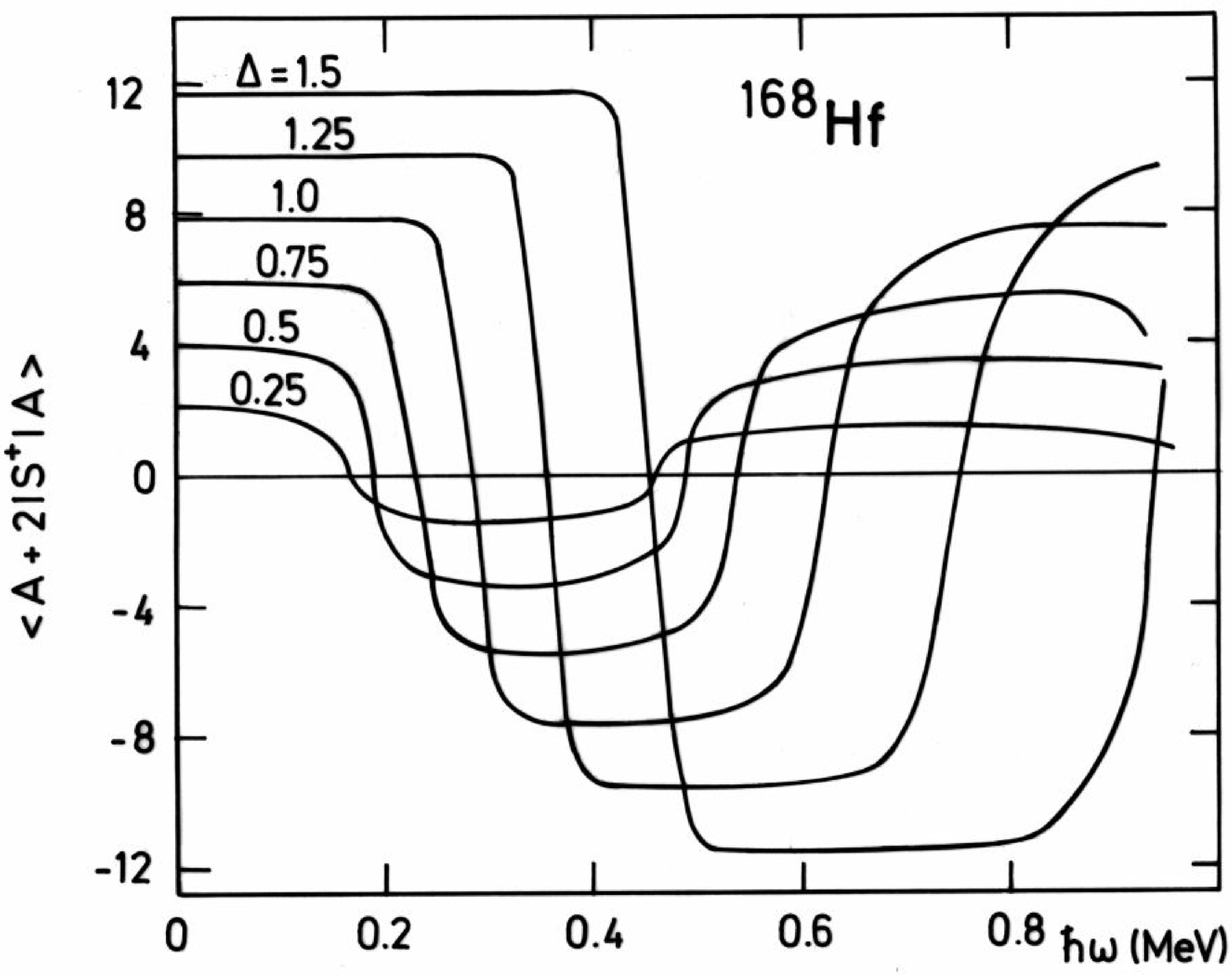,width=5.5cm}}
\caption{(Left panel) pair transfer matrix elements\cite{Nikam1986_ZPA324-241} in a prolate $j=13/2$ shell as a function of the angular velocity (in units of the deformation parameter $\kappa$). The curves labeled by $\nu$ indicate the degree to which the shell is filled. (Right panel) realistic pair transfer matrix elements\cite{Nikam1987_PLB185-269} for $^{168}$Hf as a function of the angular velocity $\omega$. Several constant values of the gap
parameters $\Delta$ are used.
\label{fig5}}
\end{figure}

In Fig.~\ref{fig5}, in the left panel, we show these spatial overlap matrix elements $\langle\varphi^{(+)}_{\nu}\vert\varphi^{(-)}_{\bar{\nu}}\rangle_\omega$ as a function of the angular velocity. For $\omega=0$ they are 1, but with increasing angular velocity they are reduced, and, in many cases, they oscillate. Such a behavior can best be understood within a one-dimensional oscillator model, in which $\pm \omega j_x$ is replaced by the pushing operator $\pm vp$. In this case the single particle wave functions can be obtained analytically as oscillator functions shifted in momentum space by $\pm v$. For the ground state these functions are Gaussians and the overlap decreases monotonically. For the first exited state we have one node in the wave function and therefore also one node in the overlap, for the second excited state there are two nodes, and this goes on accordingly for higher excited states.

It is now easy to understand the pattern of diabolic points shown in Fig.~\ref{fig4}. They are connected with vanishing matrix elements between states with positive and negative signature. For small $\Delta$, only the diagonal matrix elements $\Delta_{\nu\bar{\nu}}$ are important. The specific pattern occurs, because they are proportional to the spatial overlap matrix elements $\langle\varphi^{(+)}_{\nu}\vert\varphi^{(-)}_{\bar{\nu}}\rangle_\omega$. For increasing $\Delta$ the pair transfer is enhanced and off-diagonal elements cannot be neglected, however, the pattern of zero's does not change. This can be seen in the right panel of Fig.~\ref{fig5} where we show realistic calculations\cite{Nikam1987_PLB185-269} of the pair transfer matrix elements for the nucleus $^{168}$Hf.

\section{Berry phases in cranked HFB theory}
\label{sec5}

In his study of exceptional points Berry\cite{Berry1984_PRSocA392-45} went a step further. He investigated the topological structure of such a point and found that it is similar to that of a magnetic monopole. If in Fig.~\ref{fig3} one considers a closed path $\cal C$ with the path variable $s$ on one of the touching energy surfaces, one has, for each value of $s$, a wave function $\vert s \rangle$. These functions are eigenstates of the corresponding Hamiltonian, but their phases are open. There are certainly many ways to establish definite phase relations on this bundle
\begin{equation}
\vert\phi(s)\rangle = e^{i\gamma(s)}\vert s \rangle,
\end{equation}
but the most natural way is through parallel transport
\begin{equation}
\langle \phi(s)\vert\phi(s+\epsilon)\rangle = 1 + O(\epsilon^2).
\label{parallel-transport}
\end{equation}
The vanishing of linear terms leads to the differential equation for $\gamma(s)$
\begin{equation}
i\partial_s \gamma(s)=i\langle s\vert\partial_s\vert s\rangle~~~~~~~{\rm or}~~~~~~~
\gamma(s)=i\int_{s_0}^s\langle s^\prime\vert\partial_{s^\prime}\vert s^\prime\rangle ds^\prime
\end{equation}
Starting with the function $\phi_0=\phi(s_0)$  on the closed path $\cal C$ we find  on returning to the same point the wave function
\begin{equation}
\phi_1=e^{i\Gamma({\cal C})}\phi_0,~~~~~{\rm with}~~~~~~
\Gamma({\cal C})=i\oint_{\cal C}\langle s\vert\partial_{s}\vert s\rangle ds,
\end{equation}
which differs from the starting function $\phi_0$ by the Berry phase $\Gamma({\cal C})$. It is a characteristic non-integrable phase, a purely geometrical property of the line bundle $\{ \phi(s)\vert s\in {\cal C}\}$ which is understood in mathematical physics\cite{Simon1983_PRL51-2167} as an anholonomy and which has been observed in many areas of {physics\cite{Shapere1989}}. It can be shown that this phase is -1 for a real Hamiltonian in a two-dimensional case with two touching surfaces.

In Fig.~\ref{fig4} we apply this concept of a Berry phase to the spectrum of the rotating HFB equation (\ref{HFB-rot}). For each set of parameters ($\lambda,\omega$) we have a set of generalized Slater determinants $\Phi_n(\lambda,\omega)$. They are characterized by a specific occupation pattern of the quasiparticle levels of Fig.~\ref{fig2}. Around $\omega=\omega_c$ we observe a crossing of two bands, the ground state band ($n=0$), starting at $\omega=0$ by filling the two lowest degenerate positive quasiparticle energies $E^{(\pm)}_1$ (boldface curves in Fig.~\ref{fig2}), and an excited two-quasiparticle band ($n=1$) starting at $\omega=0$ by filling the two lowest negative quasiparticle energies $-E^{(\pm)}_1$ (thin curves in Fig.~\ref{fig2}). Since the lowest quasiparticle energy becomes negative for $\omega>\omega_1$ both bands are nearly degenerate in the region of the diabolic point.

We have to keep in mind the physical meaning of a closed a path in this surface: changes in $\lambda$ at constant $\omega$ correspond to changes in the particle number $A$, i.e. to a particle transfer ($A\rightarrow A\pm2$) for constant angular momentum $I$ and changes in $\omega$ at constant $\lambda$ correspond to changes in the angular momentum $I$ for constant particle number, i.e. to transitions with $\Delta I = \pm 2$ within the rotational band. In reality we have only integer values for $A$ and $I$ and our path is discrete, but as long as we stay within the classical limit of continuous changes in $\lambda$ and $\omega$ we encounter no problem in defining phase relations through parallel transport (\ref{parallel-transport}). Following a closed path around the diabolic point we find a Berry phase -1 for the intrinsic wave functions.

\section{Berry phases for wave functions in the laboratory frame}
\label{sec6}

So far we have considered only the topological structure of the intrinsic cranking Hamiltonian (\ref{HFB-rot}) whose eigenfunctions $\Phi(\lambda,\omega)$ are generalized Slater determinants. As discussed in section~\ref{sec3}, the cranking model fails to describe the details of a sharp level crossing properly. It is therefore by no means clear, that the topological considerations of the last section also apply to wave functions  in the laboratory frame having good particle number and angular momentum.

In order to clarify the question of Berry phases in the laboratory frame, we consider, in the following, the wave functions given in Eq.~(\ref{PSM}) making use of the projected shell model introduced by Hara and {Sun\cite{Hara1995_IJMPE4-637}}. As in the experiment, we have in this case no continuous parameters $\lambda$ and $\omega$, but only discrete integer values of the quantum numbers $A$ and $I$ which form a lattice (see Fig.~\ref{fig4}). A closed path around a possible diabolic point  on this lattice corresponds to a quantized path
\begin{equation}
\begin{array}{ccc}
(A,I+2)&\leftarrow&(A+2,I+2)\cr
\downarrow &\cdot&\uparrow\cr
(A,I)&\rightarrow& (A+2,I)
\end{array}
\label{path}
\end{equation}
Having good particle number and angular momentum the wave functions of this lattice are orthogonal and the concept of parallel transport (\ref{parallel-transport}) fails. A new concept has been introduced\cite{Sun1991_ZPA339-51} to establish phase relationships and to connect the wave functions along this path by considering the matrix elements $\langle A,I \vert \hat{C}\vert A^\prime, I^\prime\rangle$ of a suitable {\it connection operator} $\hat{C}$. A reasonable choice for this operator is $\hat{C}=\hat{Q}+\hat{S}^\dag+\hat{S}$, i.e. the quadrupole operator $\hat Q$ for states within a rotational band and the pair transfer operators $S^\dag$ and $S$ for states with constant angular momentum. This is a natural definition for the connection operator on the yrast surface of superfluid nuclei. Starting from the ground state of one nucleus, each point on this surface can be reached either by the excitation of collective rotations in gauge space (pairing rotations\cite{Bes1976_Varenna69-55}) or in $r$-space (normal rotations) and experimentally these excitations occur primarily through pair transfer or quadrupole excitations.

We then have to choose the three relative phases along our discretized path (\ref{path}) in such a fashion that all three matrix elements $\langle A,I+2 \vert \hat{Q} \vert A,I\rangle$, $\langle A,I \vert \hat{S} \vert A+2,I\rangle$, and $\langle A+2,I\vert \hat{Q} \vert A+2,I+2\rangle$ are positive. The phase of the last matrix element $\langle A+2,I+2 \vert \hat{S}^\dag \vert A,I+2\rangle$ on this closed path is then no longer open. It has to be calculated, and we have to distinguish two cases: (i) if it is positive, which is the normal case, we obviously have no Berry phase and no diabolic point within the path, (ii) if it is negative, we have a Berry phase -1. We call such a case a diabolic plaque. This is the definition of a diabolic point in the discrete case of a lattice. It obviously makes no sense to define a specific point inside the plaque. Obviously this definition of a diabolic point for the many-body wave functions with discrete quantum numbers is in complete agreement with the considerations of the last section where we found in Fig.~\ref{fig5} a sign change for the pair transfer matrix elements at the diabolic point.

\begin{figure}[th]
\centerline{\psfig{file=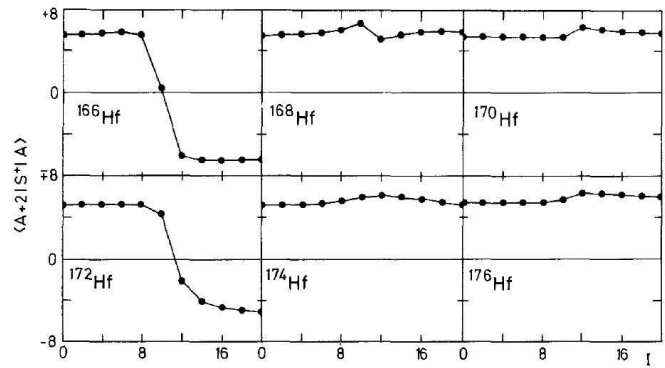,width=9.0cm}}
\caption{The HF-chain, projected shell model calculations\cite{Sun1991_ZPA339-51} for the pair transfer matrix elements $\langle A+2,I \vert \hat{S}^\dag \vert A,I\rangle$
as a function of the angular momentum $I$}
\label{fig6}
\end{figure}

In Fig.~\ref{fig6} we show realistic applications\cite{Sun1991_ZPA339-51} of the projected shell model for pair transfer matrix elements in the Hf-chain. The diabolic points in  $^{166}$Hf and $^{172}$Hf are clearly seen. These calculations are in very good agreement with mean field results in the cranking model. It shows that the Berry phase and the reduction of the pair transfer matrix elements at the diabolic point is reproduced at the mean field level rather well.

Of course the phase relations defined along the discrete path (\ref{path}) by the connection operator $\hat{C}$ can also be determined by requiring that the two transfer matrix elements $\langle A,I \vert \hat{S} \vert A+2,I\rangle$ and $\langle A,I+2 \vert \hat{S} \vert A+2,I+2\rangle$ are positive. In this case we find a sign change and zero's in the quadrupole matrix elements indicating a complete mixture of the wave functions in the area of a diabolic point.

The fact that the pair transfer matrix element changes sign in the region of the diabolic point has been called {{\it diabolic pair transfer}\cite{Nikam1987_PLB185-269,Nikam1987_PRL58-980}}. In theoretical investigations\cite{Canto1987_PLB192-4,Canto1990_PLB241-295} of Coulomb excitations, connected with pair transfer, one finds a strong reduction of the transfer cross section in this region. Of course the real proof of the existence of this change in phase would require the observation of an increase of this cross section after the diabolic region. On the other hand, in this region, there is considerable mixing and the probability for excitations within the band is also {reduced\cite{Dasso1990_PLB242-323}}. This might be the reason that, in experiment, diabolic pair transfer has, so far, not been {observed~\cite{Helmer1993_PRC48-1879}}.

\section{Concluding remarks}
\label{sec7}

Strong $ph$- and $pp$-interactions in nuclei with open proton and neutron shells lead to considerable correlations between the valence particles. One observes phase transitions
to deformed shapes and to superfluid condensates of $S$-pairs with angular momenta pointing in opposite directions. Since the Coriolis field behaves similarly to a magnetic field and tries to break the pairs and to align them along the rotational axis, one originally expected a sudden phase transition from superfluid to normal fluid connected with a strong increase of the moment of inertia. In fact, in 1971, the backbending phenomenon was found indicating such an enhancement. However, it turned out that, because of the finiteness of the system, the nucleus contains particles with rather different angular momenta. Pairs in high-$j$ intruder orbits break first. No collective pair collapse is observed, but rather a transition to an aligned two-quasiparticle band.

It turns out that these alignment processes show very interesting details. The violation of rotational symmetry by the $ph$-correlations and of gauge symmetry by the $pp$-correlations goes along with two types or rotational excitations, conventional rotations in $R_3$ and rotations in gauge {space\cite{Bes1976_Varenna69-55}}. They are characterized in a classical way through the angular velocity $\omega$ and the chemical potential $\lambda$. Therefore the intrinsic rotating HFB Hamiltonian depends on these two parameters and shows exceptional "diabolic" points in its spectrum. They  violate the no-crossing rule
and show specific topological properties resulting in non-integrable Berry phases. They follow a very specific pattern in the ($\lambda,\omega$)-plane. The origin of these diabolic points can be understood from the details of the alignment process in the rotating field. One finds, for the spatial overlap integral between the two single particle wavefunctions, oscillations, as a function of the angular velocity. They vary with the degree to which the intruder shell is filled in the nucleus and show a very specific pattern, which does not depend on the size of the pairing correlations, but only on the property of the intruder level.

The fact, that the alignment process itself is not properly described in details in this semiclassical intrinsic picture of the cranking model, and that a proper treatment needs angular momentum projection, has no influence on the pattern of diabolic points and its consequences as for instance the sign change of the interaction matrix element between the two bands as well as the sign change of the pair transfer matrix elements and its strong reduction in size in this region (diabolic pair transfer). All these phenomena also occur in the shell model, where the symmetries are treated properly, and where one deals with wave functions in the laboratory system.

On the experimental side the pattern of diabolic points has been observed. It leads to the oscillating behavior of backbending. On the other hand, so far, the strong reduction of  pair transfer (diabolic pair transfer) as well as quadrupole transition matrix elements in the region of the diabolic points has not been observed uniquely in experiment: The strong reduction of the transition probabilities has, so far, not allowed one to populate these bands beyond the diabolic region~\cite{Helmer1993_PRC48-1879} and to measure the subsequent increase of these transition probabilities. However, the fact, that one has not been able to populate these states until now is already a strong hint, that diabolic pair transfer exists.

\section*{Acknowledgements}
I want to express my gratitude to R. R. Hilton for a careful reading of the manuscript and many helpful suggestions.

\end{document}